\documentstyle[aps, prl, floats, capt15]{revtex}
\include{epsf}
\newcommand{\sruo}{Sr$_2$RuO$_4$}
\newcommand{\etal}{\ {\it et al.\/}}
\newcommand{\jpsj}{J.\ Phys.\ Soc.\ Jpn.\ }
\newcommand{\updown}{{\uparrow\downarrow}}

\title{Detailed Topography of the Fermi Surface of \sruo}

\author{C.~Bergemann,$^1$ S.~R.~Julian,$^1$ A.~P.~Mackenzie,$^2$
  S.~NishiZaki,$^3$ and Y.~Maeno$^{3,4}$} 

\address{$^1$Cavendish Laboratory, University of Cambridge, Madingley
  Road, Cambridge~CB3~0HE, United~Kingdom}

\address{$^2$School of Physics and Astronomy, University of
  Birmingham, Edgbaston, Birmingham~B15~2TT, United~Kingdom}
  
\address{$^3$Department of Physics, Graduate School of Science, Kyoto
  University, Kyoto 606-8502, Japan}

\address{$^4$CREST, Japan Science and Technology Corporation,
  Kawaguchi, Saitama 332-0012, Japan}

\date{Submitted to Physical Review Letters on 3 September 1999}

\begin{document}
\draft

\twocolumn[\hsize\textwidth\columnwidth\hsize\csname@twocolumnfalse\endcsname

\maketitle

\begin{abstract}
  
  We apply a novel analysis of the field and angle dependence of the
  quantum-oscillatory amplitudes in the unconventional superconductor
  Sr$_2$RuO$_4$ to map its Fermi surface (FS) in unprecedented detail,
  and to obtain previously inaccessible information on the band
  dispersion. The three quasi-2D FS sheets not only exhibit very
  diverse magnitudes of warping, but also entirely different dominant
  warping {\em symmetries.\/} We use the data to reassess recent
  results on $c$-axis transport phenomena.

\end{abstract}
\vspace{.5cm}]

The layered perovskite oxide \sruo\ has attracted considerable
experimental and theoretical attention since the discovery of
superconductivity in this compound five years ago \cite{maeno}. Fermi
liquid behaviour of several bulk transport and thermodynamic
properties was observed in early work \cite{maeno,maeno97}, and the
existence of mass enhanced fermionic quasiparticles was demonstrated
explicitly by the observation of quantum oscillations in the
magnetization (de Haas-van Alphen or dHvA effect) and resistivity
\cite{sruoQuosc}. Quantitative similarities between the Fermi liquid
in $^3$He and that in \sruo\ hint at the possibility of p-wave
superconducting pairing \cite{rice}. Evidence supporting such a
scenario has come from the existence of a very strong impurity effect
\cite{rkwh}, a temperature independent Knight shift into the
superconducting state \cite{ishida2}, a muon spin rotation study
indicating broken time reversal symmetry\cite{luke} and a number of
other experiments \cite{ishida,jin}. Taken together, these favour spin
triplet superconductivity with a p-wave vector order parameter and a
nodeless energy gap.

\sruo\ appears to be an ideal material in which to investigate
unconventional superconductivity in real depth: of all known compounds
exhibiting this phenomenon, \sruo\ offers the best prospects of a
complete understanding of the normal state properties within standard
Fermi liquid theory \cite{julian}. This would provide a solid
foundation for all theoretical models.  It has become clear, however,
that further progress will require very detailed knowledge of the
electronic structure of \sruo.  For example, one of the most
successful current theories relies on the assumption that the FS
consists of two sheets derived from bands with strong Ru d$_{xz,yz}$
character and one with strong Ru d$_{xy}$ character \cite{oguchi}.
These are supposed to form weakly coupled subsystems with very
different pairing interactions \cite{agterberg}. This theory has
successfully predicted non-hexagonal vortex lattice
structures \cite{agterberg2,riseman}, but it is less clear whether, in
its simplest form, it will provide a satisfactory explanation for
recent measurements of the temperature dependence of the density of
normal excitations in the superconducting state \cite{zaki}.

Of central importance to the understanding of \sruo\ is the origin of
the quasiparticle mass enhancement and how it relates to magnetic
fluctuations. Recent observations of cyclotron resonances\cite{hill}
give the promise of separating the various contributions to the
enhancement, but identifying the type of resonance and the extent to
which electron interactions are affecting the observed masses requires
more detailed knowledge of the Fermi surface than has been available
to date. Clues to the magnetic fluctuation spectrum have come from
nuclear magnetic resonance\cite{imai} and neutron scattering
experiments\cite{braden} and from calculations \cite{mazin}. Both
ferro- and antiferromagnetic fluctuations appear to be present, the
latter due to nesting of the FS.  Angular dependent magnetoresistance
oscillations (AMRO) can give information about the in-plane topography
of the FS and the extent to which it is nested. However, with three
bands crossing the Fermi level, the interpretation of AMRO data on
\sruo \cite{ohmichi} has been somewhat ambiguous.

Progress on all the issues discussed above requires high resolution,
sheet-by-sheet knowledge of the FS of \sruo.  As shown in previous
studies \cite{sruoQuosc}, the dHvA effect is ideally suited to this, as
data from individual FS sheets can be identified without ambiguity.
For this reason, we have performed a comprehensive angular dHvA study
in \sruo.  Full analysis of the data required extension and
generalisation of previously reported theoretical
treatments\cite{yamaji} of dHvA in nearly 2D materials.  As a result,
we present an unprecedentedly detailed picture of the warping of each
FS sheet.  A series of recent measurements of $c$-axis transport
phenomena are discussed in light of the full experimentally determined
dispersion.

Quantum-oscillatory effects in a crystal arise from the quantization
of the cyclotron motion of the charge carriers in a magnetic field
{\boldmath$B$}. For three-dimensional metals, only the extremal
cyclotron orbits in {\boldmath$k$}-space lead to a macroscopic
magnetization, and the quantitative treatment has been known for
decades \cite{shoenberg}. 

\begin{figure*}[t] \epsfxsize=18cm 
    \centerline{\epsfbox{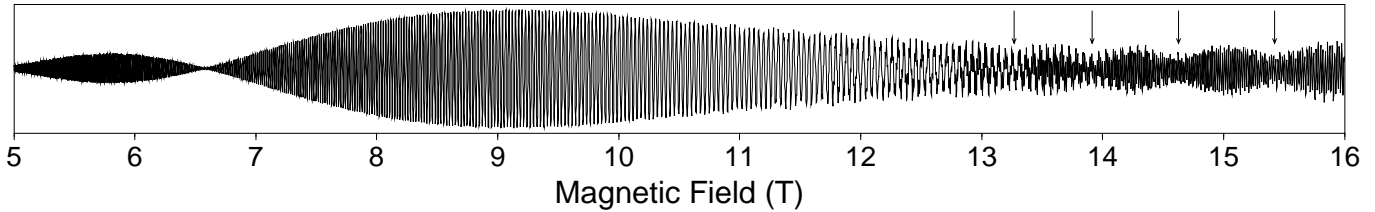}\\[1ex]} 
    \caption{Example of a dHvA field sweep on \sruo, for $\theta = 9.8^\circ$
      off the $c$-axis in the $[001] \to [110]$ rotation study. The
      vertical axis is the pick-up signal (in arbitrary units) at the
      second harmonic of the excitation frequency. At high fields,
      beats in the $\beta$-oscillations are visible, as indicated by
      arrows.}
    \label{samplesweep}
\end{figure*}

For a quasi-2D metal, the FS consists of weakly corrugated cylinders.
While such weak distortions have little effect on the cross-sectional
areas which determine the dHvA frequency, they still affect the
interference of the magnetization contributions of different parts of
the FS and therefore lead to a characteristic amplitude reduction.
Conversely, as we will show, analysis of the experimental dHvA
amplitude behaviour can reveal fine details of the topography of the
underlying Fermi cylinders.

In the most basic case, a simple corrugation of the Fermi cylinder
leads to a beating pattern in the magnetization. Analysis of the beats
for on-axis fields gives some information about the {\em magnitude\/}
of the warping, but further conclusions have to rely on assumptions
about the precise form of the corrugation \cite{sruoQuosc}. At the
next level of approximation, the FS dispersion can be determined
within the traditional scope of the Yamaji effect
\cite{yamaji,wosnitza}. A preliminary attempt to extract information
on \sruo\ in this way, however, has not achieved agreement between the
data and the predictions of the simple model \cite{yoshida}. We show
that a much more extensive treatment is needed that (a) considers
Fermi cylinder corrugation of arbitrary shape and (b) if necessary,
goes beyond the extremal orbit approximation. Also, to extract
meaningful information from experiments, one needs data of much higher
quality than has been available to date.

In the following, we will briefly present the results of a full
quantitative treatment of the oscillatory magnetization for a Fermi
cylinder that is warped arbitrarily but still compliant with the
Brillouin zone (BZ) symmetry of \sruo; details will be presented
elsewhere \cite{bergemann}. It is convenient to parameterize the
corrugation of the cylinder through an expansion of the local Fermi
wavevector,
\begin{equation}
  \label{eq:warpexpand}
    k_F(\phi,\,\kappa) = \!\!
    \sum_{\begin{array}{c} \\[-3.5ex] {\scriptstyle \mu,\,\nu\ge 0} \\[-0.8ex]
    {\scriptstyle \mu\ {\rm even}} \end{array}}
      k_{\mu\nu} \cos \nu\kappa \,
      \left\{ \begin{array}{ll} \cos \mu\phi & (\mu \bmod 4 \equiv 0) \\
          \sin \mu\phi & (\mu \bmod 4 \equiv 2)
          \end{array} \right.
\end{equation}
(see Table~\ref{warptable} for illustration).  Here, $\kappa = ck_z/2$
where $c$ is the height of the body-centered tetragonal unit cell, and
$\phi$ is the azimuthal angle of {\boldmath$k$} in the
$(k_x,\,k_y)$-plane. The $\beta$- and $\gamma$-cylinders are centered
in the BZ; symmetry allows nonzero $k_{\mu\nu}$ only
for $\mu$ divisible by 4. The $\alpha$-cylinder runs along the corners
of the BZ and has nonzero $k_{\mu\nu}$ only for $\nu$ even and $\mu$
divisible by 4, or for $\nu$ odd and $\mu \bmod 4 \equiv 2$. The
average Fermi wavevector is given by $k_{00}$, which is assumed to be
much larger than the higher-order $k_{\mu\nu}$.

One also has to consider the effect of the magnetic field:
spin-splitting drives the spin-up and spin-down surfaces apart; the
parameters $k_{\mu\nu}$ can be taken to expand weakly and linearly in
the field as in $k^\updown_{\mu\nu} = k_{\mu\nu} \pm \chi_{\mu\nu}B$.
Ordinary spin-splitting is described by $\chi_{00}$ alone, while
higher-order contributions would correspond to the underlying
electronic band structure being flatter at some points on the Fermi
surface than at others. Indeed, this anomalous spin-splitting will
prove essential for describing the $\alpha$-sheet in \sruo.

If a magnetic field is applied at polar and azimuthal angles
$\theta$ and $\phi$, the Fermi surface cross-sectional area
perpendicular to the field which cuts the cylinder axis at $\kappa$ is
given by the Bessel function expansion
\begin{eqnarray}
  \label{eq:akappa}
  a^\updown(\kappa) & = & \frac{2\pi k_{00}}{\cos\theta} 
  \sum_{\begin{array}{c} \\[-3.5ex] {\scriptstyle \mu,\,\nu\ge 0} \\[-0.8ex]
      {\scriptstyle \mu\ {\rm even}} \end{array}}
  (k_{\mu\nu} \pm \chi_{\mu\nu}B)\\
  && \hspace{-7ex} {} \times 
  J_{\mu} (\nu\kappa_F \tan\theta)
  \,\cos \nu\kappa \,
  \left\{ \begin{array}{ll} \cos \mu\phi & (\mu \bmod 4 \equiv 0) \\
      - \sin \mu\phi & (\mu \bmod 4 \equiv 2)
    \end{array} \right.
  \nonumber
\end{eqnarray}
which is a generalization of Yamaji's earlier treatment \cite{yamaji};
here, $\kappa_F = ck_{00}/2$. The total quantum oscillatory
magnetization now arises from the interference of the individual
contributions of the cylinder cross-sections; at constant chemical
potential, the fundamental component of the oscillations is
given by
\begin{equation}
  \label{eq:mtilde}
  \tilde M \propto \sum_{\updown} \int_0^{2\pi} d\kappa \,
  \sin\left(\frac{\hbar a^\updown(\kappa)}{eB}\right) \:.
\end{equation}

Eqs. (\ref{eq:akappa}) and (\ref{eq:mtilde}) describe oscillations at
the undistorted frequency $\hbar k_{00}^2 / 2e\cos\theta$, with a
characteristic amplitude modulation induced by the interference. One
can thus infer the warping parameters $k_{\mu\nu}$ by modeling
experimentally obtained amplitude data \cite{signchange}.

We have performed a thorough dHvA rotation study in the $[001] \to
[110]$ plane on a high-quality crystal of \sruo\ with $T_c >
1.3\,$K\@.  The experiments were carried out on a low noise
superconducting magnet system in field sweeps from 16\,T to 5\,T, at
temperatures of 50\,mK\@. A~modulation field of 5.4\,mT amplitude was
applied to the grounded sample, and the second harmonic of the voltage
induced at a pick-up coil around the sample was recorded, essentially
measuring $\partial^2M/\partial B^2$, with well-established extra
contributions from the field modulation, impurity scattering, and
thermal smearing \cite{shoenberg}.

\begin{figure*}[t] \parbox{1.44\columnwidth}{\epsfxsize=1.44\columnwidth 
    \epsfbox{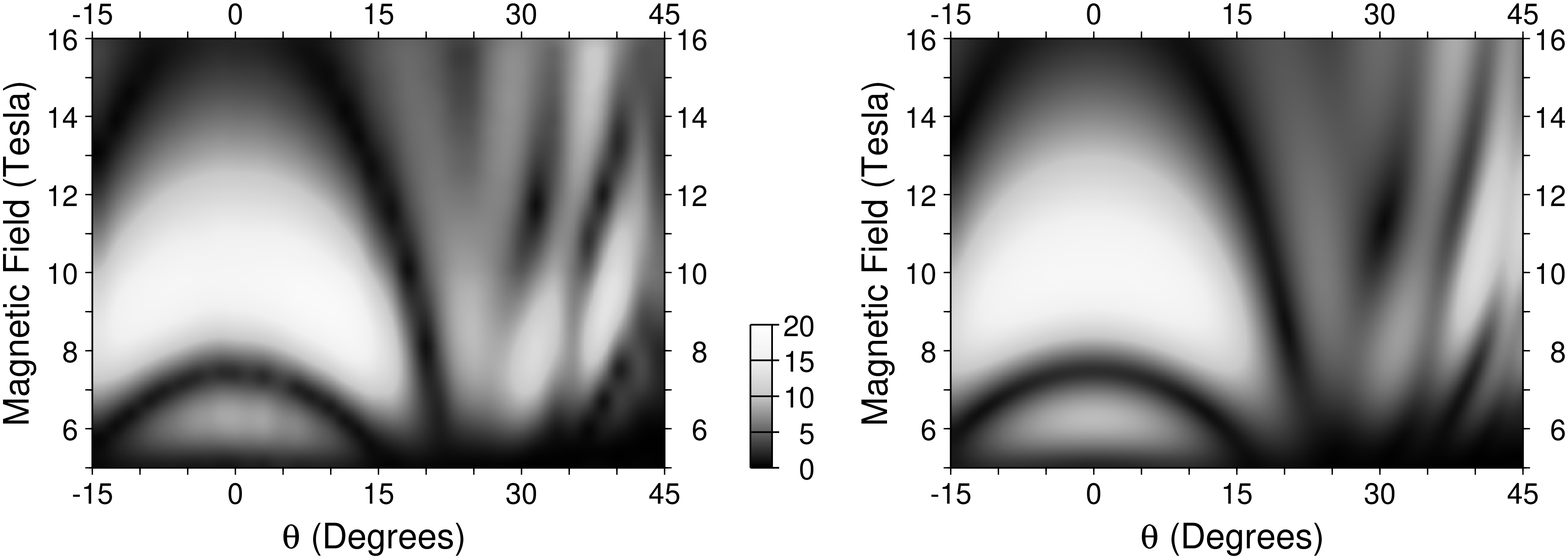}} \captiononefive{Density plot of the
    dHvA amplitude (in arbitrary units) of the $\alpha$-frequency in
    \sruo, in the experimental $[001] \to [110]$ rotation study
    (left), and comparison with the theoretical simulation, using the
    warping parameters in Table~\ref{warptable} (right). The
    theoretical calculation incorporates experimental effects such as
    the field modulation amplitude characteristic.}
    \label{PRL_alpha}
\end{figure*}

A~typical signal trace, demonstrating the high quality of our data,
can be viewed in Fig.~\ref{samplesweep}.  Overall, 35 \nopagebreak
such sweeps were performed, at angular intervals of $2^\circ$. For
each of the dHvA frequencies corresponding to the three FS sheets, we
have extracted the local dHvA amplitude through filtering in the
Fourier domain. The dHvA amplitude can then be visualized versus
magnitude and direction of the field, as shown in the density plot in
Fig.~\ref{PRL_alpha} for the $\alpha$-sheet.  We have also performed
similar data analysis for a second extensive rotation study of short
high-field ($18\,{\rm T}$ to $15\,{\rm T}$) sweeps in the $[001] \to
[100]$ plane; these runs used a different sample.  We now turn to the
results of the analysis for all three FS sheets, where
Table~\ref{warptable} presents the deduced values for the
$k_{\mu\nu}$.

\begin{figure}[t] \epsfxsize=8.7cm 
    \centerline{\epsfbox{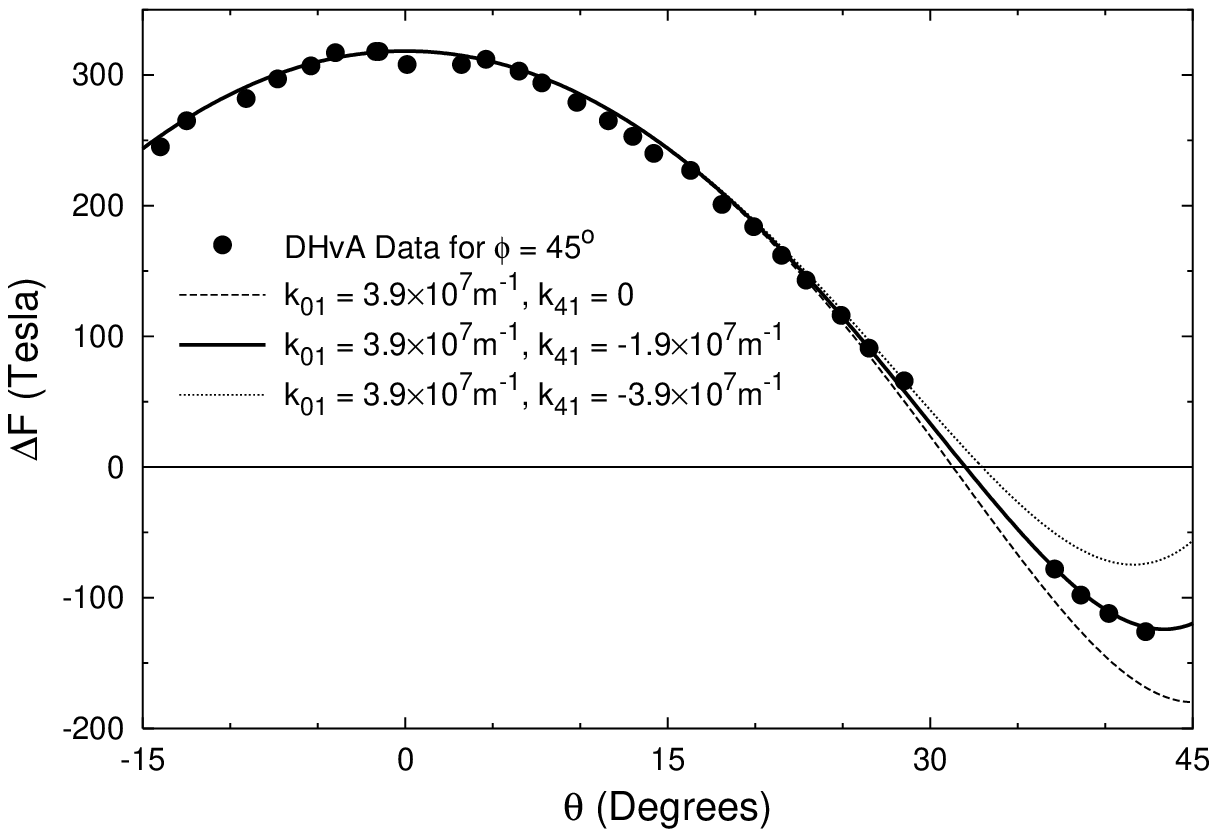}\\[1ex]} 
    \caption{Angular dependence of the beat frequency of the 
      $\beta$-oscillations in \sruo\ (cf.\
      Fig.~\ref{samplesweep}) in the $[001] \to [110]$ rotation study
      (full circles).  $\Delta F(0^\circ)$ fixes $k_{01}$ at
      $3.9\times 10^7\,{\rm m}^{-1}$; the data are compared against
      the predicted behaviour for cylindrically symmetric warping
      (dashed line), for zero dispersion along the BZ axes (dotted
      line), and for $k_{04}$ as tabulated in Table~\ref{warptable}
      (solid line).}
    \label{betadifffreq_PRL}
\end{figure}

{\em $\alpha$-Sheet} --- The most striking features of the data in
Fig.~\ref{PRL_alpha} are qualitative differences with similar data in
the $[001] \to [100]$ plane (Ref.~\onlinecite{yoshida} and the present
study), and the absence of spin-zeroes that should be visible as
vertical black lines.  We are able to account for both effects, and
produce near-perfect agreement with experiments as seen by comparing
the two panels of Fig.~\ref{PRL_alpha}, using the $k_{\mu\nu}$ as given
in Table~\ref{warptable}. The dominant $k_{21}$-term affects the dHvA
amplitude for $\phi = 45^\circ$ but has no effect for fields in the
$[001] \to [100]$ plane. The absence of spin-zeroes arises from the
presence of a finite $\chi_{21} \simeq -5\times 10^5\,{\rm m}^{-1}{\rm
  T}^{-1}$ in addition to $\chi_{00} = 10.4\times 10^5\,{\rm
  m}^{-1}{\rm T}^{-1}$. The same parameters account
equally well for data from $[001] \to [100]$ rotations and for the
amplitude of the second harmonic for both rotation directions. It
should be noted that the warping of the $\alpha$-cylinder is so weak
--- at some angles and fields it is smaller than the Landau level
spacing --- that the success of the model requires the use of our
exact treatment beyond the extremal orbit approximation.

{\em $\beta$-Sheet} --- The warping is comparatively large, and the
extremal orbit approximation is valid over most of the angular range.
The Yamaji angle of $32^\circ$ and the variation of the (relatively
fast) beating frequency $\Delta F$ with $\theta$
(Fig.~\ref{betadifffreq_PRL}) reveal that the dominant warping
parameters are $k_{01}$ and $k_{41}$, as tabulated in
Table~\ref{warptable}. They have opposite signs, so the $c$-axis
dispersion is {\em largest along the zone diagonals.\/} It is
difficult to extract meaningful information on the higher-order terms,
but we believe that the data set an upper bound for a double warping
contribution of $|k_{02}| < 10^7\,{\rm m}^{-1}$. The spin-splitting
behaviour is intricate, and while it is certain that higher-order
$\chi_{\mu\nu}$ are needed to explain the data, it is impossible at
this stage to extract them without ambiguity.

{\em $\gamma$-Sheet} --- Due to stronger impurity damping, the
$\gamma$-signal is only observable at fields of more than 13\,T\@.
Along $[001] \to [110]$, its amplitude peaks at $\theta = \pm
13.7^\circ$, implying that the dominant corrugation is {\em double\/}
warping, i.e. $k_{02} \gg k_{01}$. We obtained a rough estimate of its
strength from the {\em sharpness\/} of this amplitude maximum --- it
is difficult to assess $k_{02}$ from an on-axis beating pattern, as
that cannot be established in the short field range over which
$\gamma$-oscillations are visible. For the $[001] \to [100]$ rotation,
the amplitude maximum occurs at $\theta = 14.6^\circ$. The
deviation of the two measured $\theta$-values from each other and from
the simple Yamaji predicition of $14.1^\circ$ yields $k_{42}$, whose
sign implies that the $c$-axis dispersion is {\em largest along the
  zone axes.\/} At present, it is not possible to extract reliable
information on the spin-splitting parameters.

To the order of expansion given here, it is possible to calculate the
contribution of each FS to the $c$-axis conductivity. The
$\beta$-sheet dominates with a 86\% share, compared to 8\% for the
$\alpha$- and 6\% for the $\gamma$-sheet. This provides
new insight into recent AMRO experiments by Ohmichi\etal\cite{ohmichi}
as it strongly suggests that the AMRO signal originates predominantly
from the $\beta$-sheet (rather than the $\alpha$-sheet as had
previously been assumed). The AMRO data then fix the ``squareness''
parameter of the $\beta$-cylinder as $k_{40} = 5.3\times10^8\,{\rm
  m}^{-1}$, giving quantitative information about the FS nesting on
that sheet.

Our work also helps to clarify the interpretation of absorption
spectra in cyclotron resonance \cite{hill}, which have recently been
used to assess quasiparticle mass enhancements in \sruo. The
``periodic orbit resonance'' geometry used in those experiments
assesses modulations in the $c$-axis Fermi velocity. The strongest
signals from such resonances in their simplest form should again stem
from the $\beta$-sheet, and the unusual warping symmetry would lead to
dominant resonances at $4\omega_c,\,8\omega_c,\,\ldots$ for the
$\beta$- and $\gamma$-orbits, and at $2\omega_c,\,4\omega_c,\,\ldots$
for the $\alpha$-orbit.

High-precision dHvA itself can provide normally inaccesible
information on spin-dependent many-body effects, by measuring
both the specific heat and the spin susceptibility mass enhancements.
For the $\alpha$-sheet, we have $m^\star/m = 3.4$ and $m^\star_{\rm
  susc}/m = 4.1$, the latter obtained from spin-splitting analysis
assuming $g \simeq 2$.  We would expect the ratio $m^\star_{\rm
  susc}/m^\star$ to diverge at a ferromagnetic quantum critical point;
the small ratio here suggests that, at least for the $\alpha$-sheet,
the paramagnetic susceptibility enhancement is matched by specific
heat contributions from phonons or large-$q$ spin fluctuations
\cite{julian,braden}.

\addtolength{\textfloatsep}{-3mm}

\begin{table}[t]
  \caption{Summary of the extracted warping parameters $k_{\mu\nu}$
    (in units of $10^7$\,m$^{-1}$) of the three
    FS sheets of \sruo \protect\cite{signchange}. Entries
    symbolized by a long dash are forbidden by 
    the BZ symmetry. The first row visualizes
    the warping for the different values of $\mu$ and $\nu$, 
    on top of a large $k_{00}$.}
  \label{warptable}

  \begin{tabular}{ccccccc}\\[-.2em]
    \multicolumn{2}{l}{
    \epsfxsize=1.2cm\epsfbox{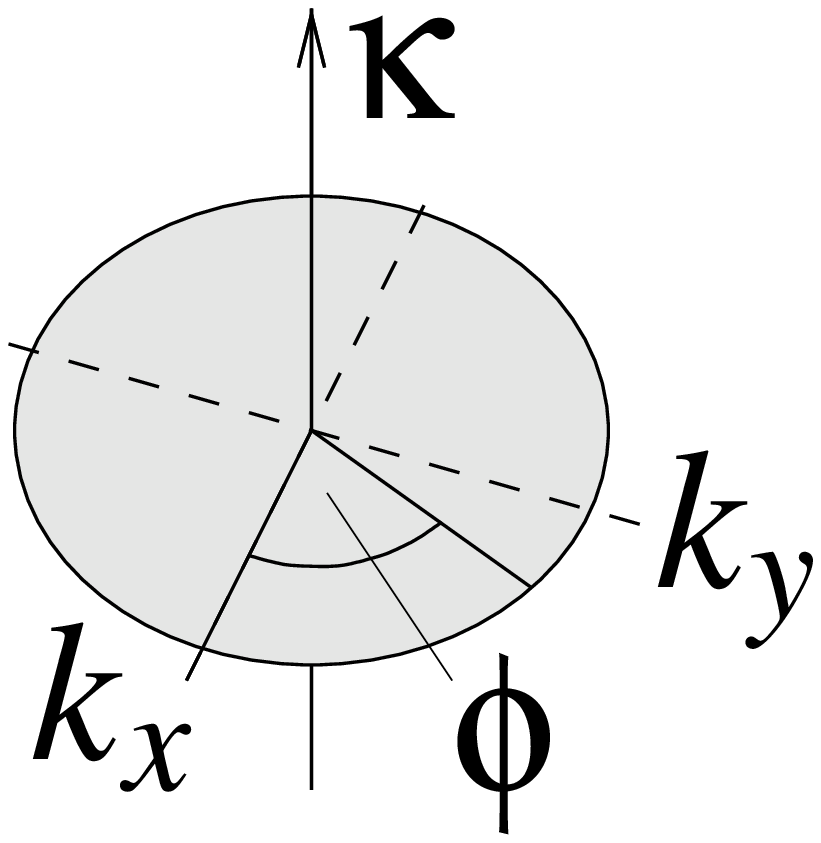}} &
    \epsfxsize=1.2cm\epsfbox{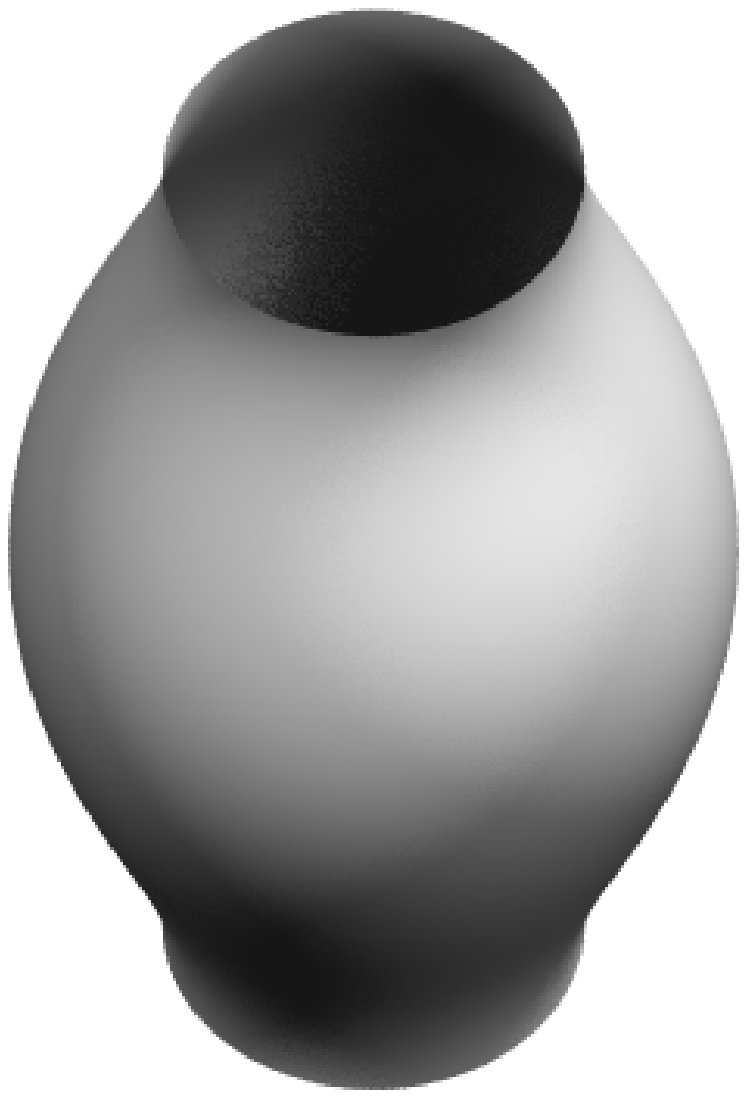} & 
    \epsfxsize=1.2cm\epsfbox{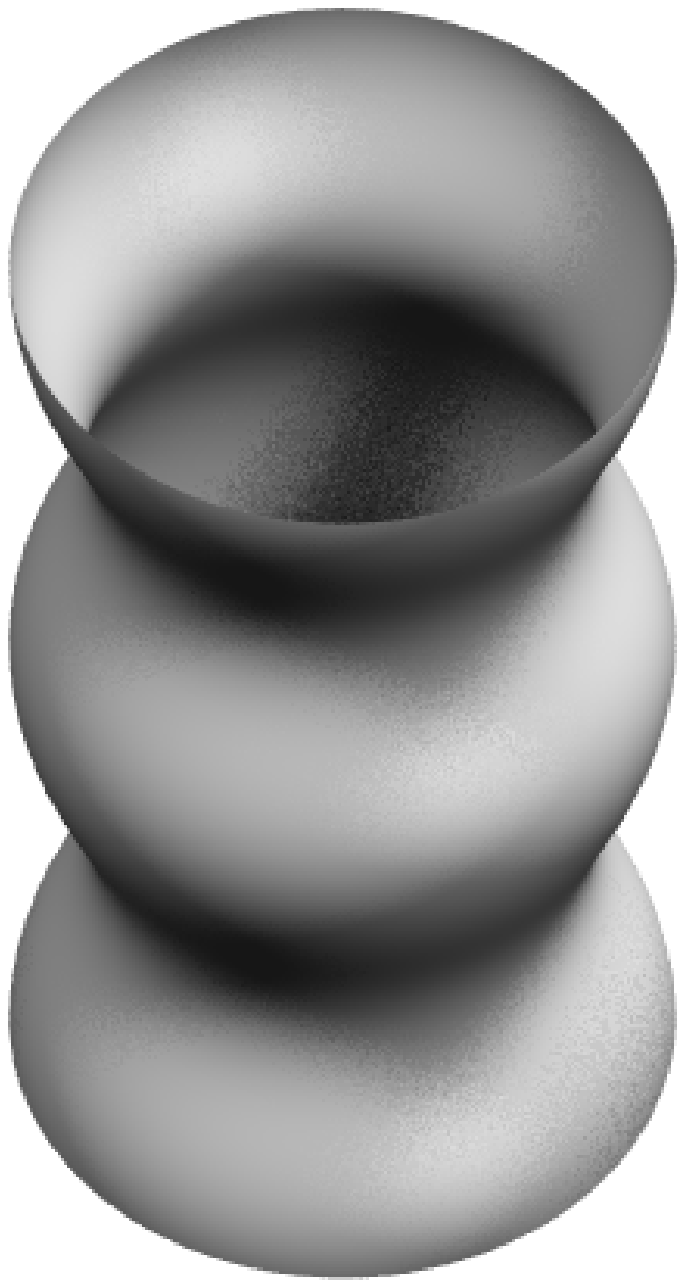} & 
    \epsfxsize=1.2cm\epsfbox{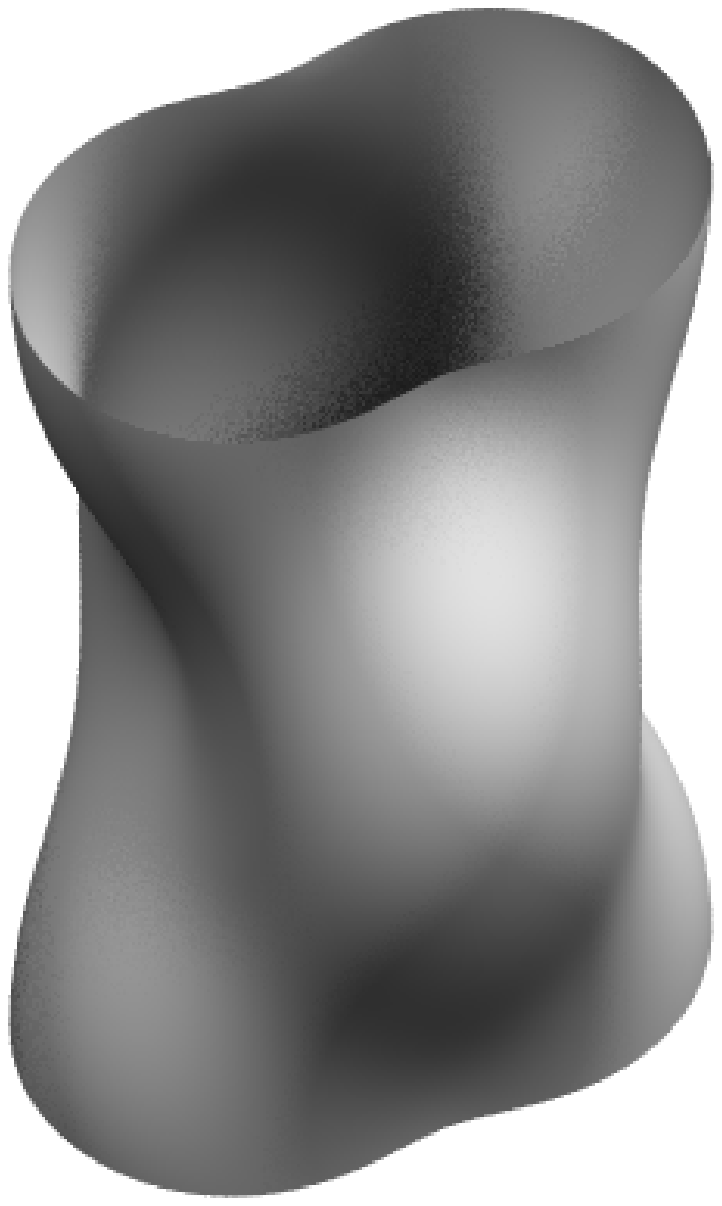} & 
    \epsfxsize=1.2cm\epsfbox{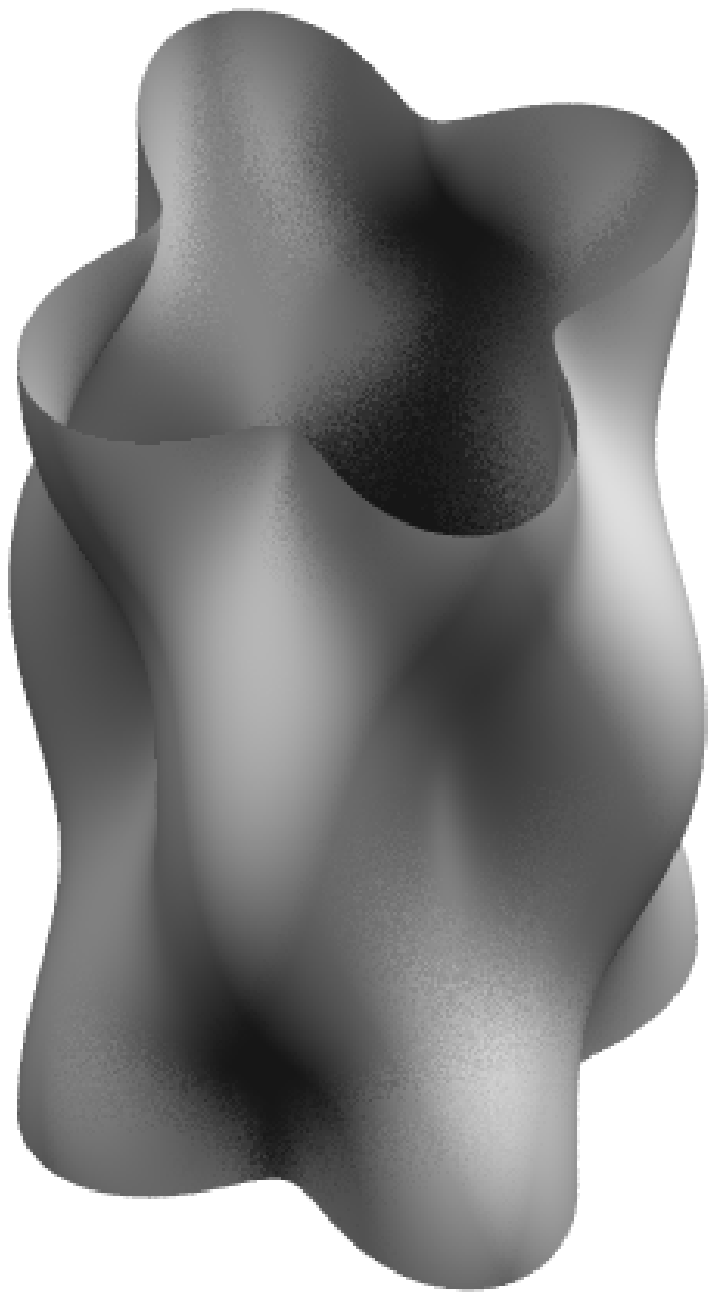} & 
    \epsfxsize=1.2cm\epsfbox{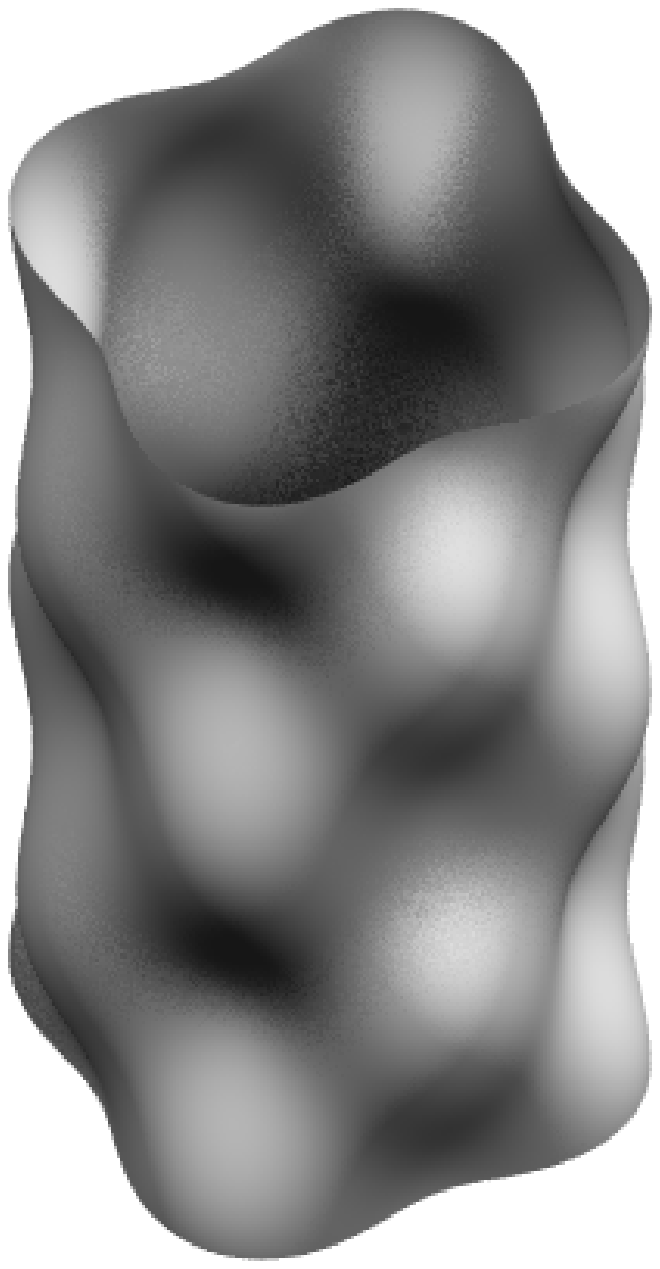} \\[1ex]
    & $k_{00}$ & $k_{01}$ & $k_{02}$ & $k_{21}$ & $k_{41}$ & $k_{42}$ \\\hline
    $\alpha$ & 304.7 & ---   & 0.33 & 1.3   & ---    & $-1.0$ \\
    $\beta$  & 623 & 3.9   & small & ---  & $-1.9$ & small   \\
    $\gamma$ & 754 & small & 0.5  & ---  & small  & 0.3    \\
  \end{tabular}
  
  \vspace{1ex}

\end{table}

Finally, an intriguing feature of the present study is the qualitative
difference between the dominant warping symmetries observed for the
three FS sheets of \sruo. Detailed comparison with the results of band
structure calculations would be very informative to test the accuracy
of these computations for the weak out-of-plane dispersions in
quasi-2D metals in general. For \sruo\ in particular, this would also
be a check on the commonly made assignment of the Ru d$_{xz,yz}$
orbital character to the $\alpha$- and $\beta$-sheets, and the
d$_{xy}$ character to the $\gamma$-sheet, which lies at the foundation
of theories of orbital dependent superconductivity.

In summary, the full analysis of angle-dependent dHvA amplitude data
has emerged as an extremely powerful tool to determine the exact
topography of the FS in quasi-2D metals. We have been able to extract
quantitative information on the corrugation of all three Fermi
cylinders of \sruo. The single warping of the $\beta$-sheet provides
most of the $c$-axis dispersion, while the ellipsoidal warping of the
$\alpha$- and the double warping of the $\gamma$-sheet are less
significant.

We wish to thank S.~Hill, E.~M.~Forgan, A.~J.~Schofield,
G.~J.~McMullan, and G.~G.~Lonzarich for stimulating and fruitful
discussions. The work was partly funded by the U.K. EPSRC\@. C.B.
acknowledges the financial support of Trinity College, Cambridge; and
A.P.M.  gratefully acknowledges the support of the Royal Society.


\begin{references}
\bibitem{signchange} The dHvA effect is insensitive to $k_{\mu 0}$ for
  $\mu > 0$; these parameters can only be probed by other methods such
  as AMRO.  Ambiguities can also arise for the {\em sign\/} of the
  $k_{\mu\nu}$: while the average Fermi wavevector $k_{00}$ and the
  spin-splitting $\chi_{00}$ must be positive, the dHvA amplitude is
  invariant under a simultaneous sign change of the other $k_{\mu\nu}$
  and $\chi_{\mu\nu}$ for either all $\nu$, or all even $\nu$, or all
  odd $\nu$.
\end{references}
\end{document}